\documentclass[twocolumn,pre,showpacs,amsmath,amssymb,floatfix]{revtex4}
\usepackage{graphicx} 
\usepackage{dcolumn}  
\usepackage{bm}       
\begin{document}

\preprint{Draft}

\title{Calorimetric study of the nematic to smectic-\textit{A} and
smectic-\textit{A} to smectic-\textit{C} phase transitions in
liquid-crystal+aerosil dispersions}

\author{A. Roshi and G. S. Iannacchione}
 \email{gsiannac@wpi.edu}
\affiliation{Department of Physics, Worcester Polytechnic
Institute, Worcester, Massachusetts 01609, USA.}

\author{P. S. Clegg}
\affiliation{Department of Physics and Astronomy, University of
Edinburgh, Edinburgh EH9 3JZ, United Kingdom.}

\author{R. J. Birgeneau}
\affiliation{Department of Physics, University of California,
Berkeley, California, 94720, USA.}

\author{M. E. Neubert}
\affiliation{Liquid Crystal Institute, Kent State University,
Kent, Ohio 44242, USA.}

\date{\today}


\begin{abstract}
A high-resolution calorimetric study has been carried out on
nano-colloidal dispersions of aerosils in the liquid crystal
4-\textit{n}-pentylphenylthiol-4'-\textit{n}-octyloxybenzoate
($\bar{8}$S5) as a function of aerosil concentration and
temperature spanning the smectic-\textit{C} to nematic phases.
Over this temperature range, this liquid crystal possesses two
continuous XY phase transitions: a fluctuation dominated nematic
to smectic-\textit{A} transition with $\alpha \approx \alpha_{XY}
= -0.013$ and a mean-field smectic-\textit{A} to
smectic-\textit{C} transition. The effective critical character of
the \textit{N}-Sm\textit{A} transition remains unchanged over the
entire range of introduced quenched random disorder while the peak
height and enthalpy can be well described by considering a cut-off
length scale to the quasi-critical fluctuations. The robust nature
of the \textit{N}-Sm\textit{A} transition in this system contrasts
with cyanobiphenyl-aerosil systems and may be due to the mesogens
being non-polar and having a long nematic range. The character of
the Sm\textit{A}-Sm\textit{C} transition changes gradually with
increasing disorder but remains mean-field-like. The heat capacity
maximum at the Sm\textit{A}-Sm\textit{C} transition scales as
$\rho_S^{-0.5}$ with an apparent evolution from tricritical to a
simple mean-field step behavior. These results may be generally
understood as a stiffening of the liquid crystal (both the nematic
elasticity as well as the smectic layer compression modulus $B$)
with silica density.
\end{abstract}



\maketitle


\section{INTRODUCTION}
\label{sec:intro}

The study of quenched random disorder effects addresses many
fundamental issues of current interest in statistical mechanics.
Recent experimental advances have shed considerable light onto the
random-field theoretic approach, believed to be underlying the
physics of quenched random disorder (QRD). In particular, these
efforts have led to the systematic study of the random-field model
for transitions that break a continuous symmetry~\cite{Germano04}.
This model is applicable to a terrific range of phenomena: This
include unique assemblies of colloids, complex fluids, charge
density waves, spin glasses, and doped semiconductors, to name a
few. The experimental efforts to date have focussed on idealized
physical systems in order to isolate the essential features of
quenched random disorder. Considerable research has been carried
out on the superfluid transition of $^4$He and $^4$He-$^3$He
mixtures in a variety of porous media~\cite{Reppy92,Chan96} as
well as doped magnetic systems~\cite{Birgeneau98} but many
questions remain due to the quantum nature of the former and the
glass-like behavior of the latter~\cite{Germano04}.

In the random-field approach, the effect of the disorder is mapped
onto a local field, $\vec{h}(\vec{r})$, directly coupled to the
order parameter. This local field varies randomly through the
system on length scales smaller than the length scales of the
ordered phase such that $\langle \vec{h} \rangle_{\vec{r}} = 0$
while $\langle \vec{h}\cdot\vec{h} \rangle_{\vec{r}} \neq 0$.
Because the nature of the imposed disorder is modelled as a dilute
(or weak) random-field, the effect of QRD may be understood
statistically. By studying in detail good realizations of
particular universality classes, the results can be applied more
generally to a wide variety of physical systems. A particularly
fruitful physical system to explore QRD effects in general and the
random-field model in particular, have been nano-colloidal
dispersions of an aerosil gel in a liquid crystal (LC) host
(LC+sil)~\cite{Kutnjak05,Germano04}.

The most well studied LC+sil phase transition has been the
continuous nematic to smectic-\textit{A} (\textit{N}-Sm\textit{A})
phase transition. This transition involves the breaking of a
continuous symmetry and belongs, though not simply, to the 3D-XY
universality class \cite{Gennes93}. In general, high-resolution
x-ray~\cite{Park02,Leheny03,Clegg03a,Clegg03b} and
calorimetry~\cite{Zhou97b,Haga97a,Haga97b,Germano98,Marinelli01,Kutnjak05}
studies on the \textit{N}-Sm\textit{A}+sil transition have found
that the (quasi-) long-range ordered smectic phase is destroyed
for all densities of aerosil. However, below a silica density of
about $0.1$~grams of SiO$_2$ per cm$^3$ of LC (the so-called
conjugate density, $\rho_S$~\cite{Germano04}, whose units will be
dropped hereafter), a pseudo-transition displaying quasi-critical
behavior at $T = T$* persists. The scattering from the short-range
smectic order is well described by both a bulk thermal structure
factor (dominate above the pseudo-transition $T$*) and a
random-field structure factor (dominate below $T$*) given by the
bulk thermal form squared, the so-called "lorentzian+lorentzian
squared" form~\cite{Park02,Leheny03}. Although the smectic
correlation length is finite, it is large (spanning many mean-void
lengths of the silica gel) and exhibits a power-law divergence on
cooling to $T$* at which it saturates for 8CB~\cite{Park02} and
8OCB~\cite{Clegg03a} in aerosils or begins to decrease upon
further cooling towards the Sm\textit{C} phase for $\bar{8}$S5+sil
system~\cite{Clegg03b}. This broadening observed for
$\bar{8}$S5+sil~\cite{Clegg03b} and anisotropic (field aligned)
$\bar{8}$S5+sil~\cite{Liang04} samples, could be due to a
distribution of Sm\textit{C} tilt angles and result in a lower
average order parameter squared. The calorimetry studies found
sharp, quasi-critical, power-law divergences for the heat capacity
for $\rho_S < 0.1$, while all phase transitions appear to be
elastically smeared for larger $\rho_S$~\cite{Germano04}. The
evolution of an effective critical heat capacity exponent
$\alpha_{eff}$ with aerosil density is consistent with a gradual
drop in the nematic susceptibility with increasing gel density for
all \textit{N}-Sm\textit{A} transitions studied to date.

A detailed scaling analysis of the \textit{N}-Sm\textit{A}
transition combining x-ray and calorimetry results through the
quasi-critical transition for low aerosil density revealed the
importance of both random-field and finite-size-like
effects~\cite{Leheny03,Germano03}. The dimensionality of the
system also played a role and was consistent with the expected
dimensional rise, $d_{rf} = d + 2$, for a system with random-field
disorder~\cite{Aharony76,Young77}. This work also found in general
that the effective random-field strength (or variance) scales as
$\rho_S$ over the entire range of silica densities studied thus
far, although there are indications of different scalings above
and below $\rho_S = 0.1$~\cite{Germano03}. The experimental and
theoretical results to date support the view that a
random-field-XY (\textit{RF}-XY) system has no new critical point.

Recently, tilted smectic phases have become the focus of studies
on quenched random disorder effects~\cite{Clegg03b,Kutnjak05}. The
smectic-\textit{A} to smectic-\textit{C}
(Sm\textit{A}-Sm\textit{C}) phase transition involves the breaking
of a continuous symmetry and is described by two parameters, the
tilt and azimuthal angles. This transition belongs to the 3D-XY
universality class but is mean-field in character due to the
relatively long bare correlation length of the Sm\textit{C}. The
strong coupling between tilt and smectic layer compression for the
Sm\textit{C} phase appears to place the Sm\textit{A}-Sm\textit{C}
transition always close to a (Landau) mean-field tricritical
point~\cite{Gennes93}. A consequence of the tilt angle's
sensitivity to the layer elasticity is that this transition is
much more strongly disordered by perturbations that distort the
smectic layers, as in the strong disorder of aerogels, than the
\textit{N}-Sm\textit{A}~\cite{Zhou97a,Haga97c}. For weaker aerosil
gel induced disorder, the Landau mean-field heat capacity
signature was found to be unaffected by hydrophobic
aerosils~\cite{Haga97a} while a recent high-resolution x-ray
scattering study on a different LC in a hydrophilic aerosil found
little change in the tilt angle temperature dependence with
$\rho_S$ \cite{Clegg03b}. This robustness to weak disorder may be
a consequence of the mean-field character of the transition
placing it effectively at its upper critical dimension. However,
when the transition from the Sm\textit{A} phase goes to the chiral
analog of the Sm\textit{C} phase (Sm\textit{C}$^{*}$), the effect
of even the aerosil induced disorder generally suppresses and
smears the transition~\cite{Kutnjak05}.

The present work focusses on the effect of quenched random
disorder induced by a nano-colloidal dispersion of hydrophilic
type-300 aerosil forming a mass-fractal gel within the liquid
crystal
4-\textit{n}-pentylphenylthiol-4'-\textit{n}-octyloxybenzoate
($\bar{8}$S5) that closely follows the previously reported x-ray
studies~\cite{Clegg03b}. High-resolution calorimetry has been
carried out on these $\bar{8}$S5+sil dispersions as a function of
aerosil concentration and temperature spanning the
smectic-\textit{C} to nematic phases. This liquid crystal
possesses two continuous XY phase transitions of interest. The
first is a fluctuation dominated nematic to smectic-\textit{A}
characterized by a heat capacity exponent $\alpha \lesssim 0$. The
critical character of the \textit{N}-Sm\textit{A} transition
remains unchanged with the introduction of the quenched random
disorder while its enthalpy and heat capacity maximum decreases in
a manner consistent with a finite-size-like scaling without any
obvious crossover from soft to stiff gel seen on other LC+sil
systems. The stability of the heat capacity quasi-critical
behavior for this system with QRD is consistent with our
understanding of the influence of the gel on the underlying
nematic order. Since in the bulk material the nematic
susceptibility is low before Sm\textit{A} order begins to form,
the gel has little scope to reduce this quantity.

The second is a mean-field Landau tricritical smectic-\textit{A}
to smectic-\textit{C} phase transition. The
Sm\textit{A}-Sm\textit{C} remains mean-field for all aerosil
concentrations studied with an continuous evolution from the
bulk's tricritical to a simple mean-field step heat capacity
behavior analogous to that seen for the pure
Sm\textit{A}-Sm\textit{C}* transition in binary LC
mixtures~\cite{Chan93}. In particular, the
Sm\textit{A}-Sm\textit{C} heat capacity maximum at the transition
scales as $\rho_S^{-0.5}$. The stable mean-field character of the
Sm\textit{A}-Sm\textit{C} with QRD may be a consequence of this
transition being effectively, due to the long range interactions,
at its upper critical dimension. The observed crossover from
tricritical to a simple mean-field step behavior for the
Sm\textit{A}-Sm\textit{C} and the continuous diminishing of the
\textit{N}-Sm\textit{A} heat capacity peaks may be understood as a
continuous stiffening of the liquid crystal elasticity for the
nematic and smectic structure with increasing silica density.
However, many theoretical challenges are evident.

Section~\ref{sec:exp} describes the preparation of the
$\bar{8}$S5+sil dispersions as well as the ac-calorimetry
technique employed. Section~\ref{sec:results} presents the results
of the calorimetric study, while Section~\ref{sec:disc} discusses
the significance of the evolution of an XY transition as well as a
crossover from Landau tricritical to mean-field continuous
transition with increasing quenched disorder. Directions for
future study will also be discussed.

\section{SAMPLES AND CALORIMETRY}
\label{sec:exp}

\begin{figure}
\includegraphics[scale=0.5]{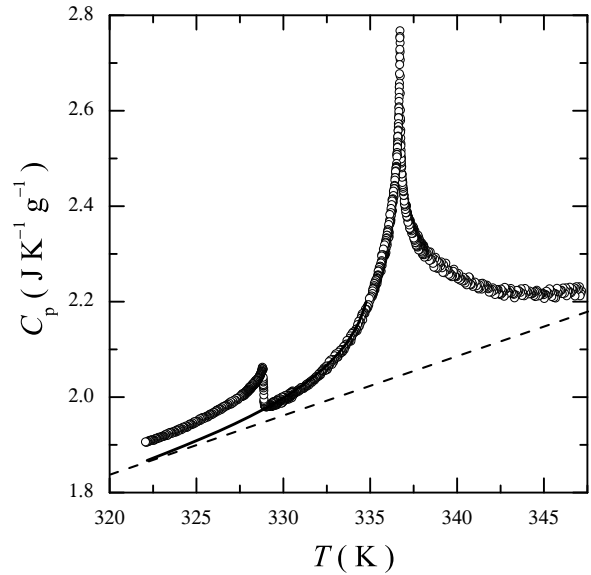}
\caption{ \label{CPvsT} The specific heat of bulk $\bar{8}$S5 on
cooling spanning the nematic, smectic-\textit{A}, and
smectic-\textit{C} phases. The dashed line represents the linear
$C_p$(background) used to extract the excess $C_p$ associated with
the \textit{N}-Sm\textit{A} transition, $\Delta C_p(NA)$. The
solid line represents the low-temperature wing of the
\textit{N}-Sm\textit{A} transition and is used to extract the
excess $C_p$ associated with the Sm\textit{A}-Sm\textit{C}
transition, $\delta C_p(AC)$. }
\end{figure}

\begin{figure}
\includegraphics[scale=0.5]{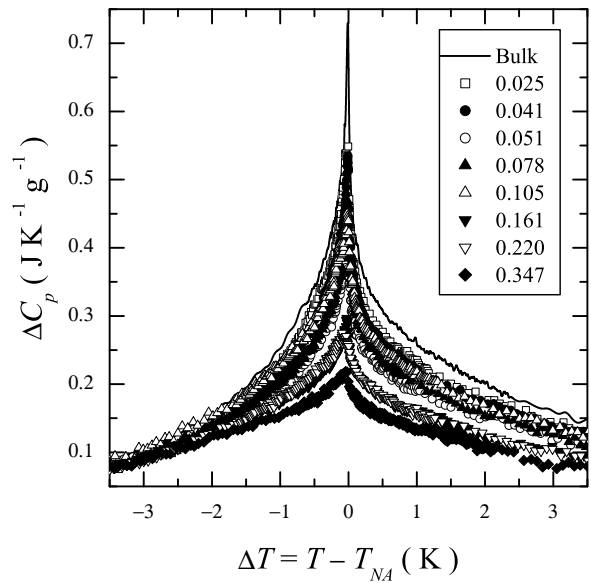}
\caption{ \label{DCPvsDTNA} Excess specific heat $\Delta C_p$ of
the \textit{N}-Sm\textit{A} transition $\pm 3.5$~K about $T_{NA}$
for bulk $\bar{8}$S5 and $\bar{8}$S5+sil samples. Inset lists the
$\rho_S$ for each data set shown. The temperature range shown
corresponds approximately to $\pm 10^{-2}$ width in reduced
temperature. }
\end{figure}

The liquid crystal $\bar{8}$S5, synthesized at Kent State
University, was used after degassing in the isotropic phase for
two hours. The best literature reported transition temperature
values in the bulk for this liquid-crystal molecule ( $M_w =
412.64$~g~mol$^{-1}$ ) are $T_{IN}^o \cong 359.6$~K for the weakly
first-order isotropic to nematic (\textit{I}-\textit{N})
transition, $T_{NA}^o \cong 336.58$~K for the XY-like continuous
nematic to smectic-\textit{A} (\textit{N}-Sm\textit{A})
transition, and $T_{AC}^o \cong 329.35$~K for the monotropic
Landau tricritical smectic-\textit{A} to smectic-\textit{C}
(Sm\textit{A}-Sm\textit{C}) transition~\cite{Schantz78}. At lower
temperatures on cooling, a monotropic smectic-\textit{C} to
crystal-\textit{B} (Sm\textit{C}-Cr\textit{B}) transition occurs
at $T_{CB}^o \sim 304$~K. The strongly first-order
Crystal-Sm\textit{A} (Cr-Sm\textit{A}) transition occurs
reproducibly on heating at $T_{CrA}^o \sim 332$~K. The measured
transitions temperatures for our bulk material occur at $T_{NA}^o
\cong 336.64$~K, $T_{AC}^o \cong 328.96$~K, and $T_{CrA}^o \sim
331.0$~K, which are in reasonable agreement with the literature
bulk values. However, $\bar{8}$S5 is known to age with its
transition temperatures continuously shift downward with time,
especially when heated into the isotropic
phase~\cite{Schantz78,Safinya81}. Fortunately, the
\textit{N}-Sm\textit{A} and Sm\textit{A}-Sm\textit{C} transitions
remain relatively sharp, well defined, and consistent in shape
during the sample aging, see Fig.~\ref{CPvsT} and \ref{DCPvsDTNA}
as well as Table~\ref{tab:Cp-sum}. Although the aging of this LC
is unavoidable, the samples studied in this work all experienced
the same preparation method and thermal history, thus the relative
evolution of these transitions with aerosil disorder should be
preserved.

\begin{table*}
\caption{ \label{tab:Cp-sum} Summary of the calorimetric results
for $\bar{8}$S5+sil samples. Shown are the conjugate silica
density ($\rho_S$ in grams of aerosil per cm$^3$ of $\bar{8}$S5),
the mean-void length $l_o = 2/a\rho_S$ (where $a =
300$~m$^2$~g$^{-1}$ is the specific surface area of this aerosil)
within the gel in \AA, the \textit{N}-Sm\textit{A} ($T_{NA}$) and
the Sm\textit{A}-Sm\textit{C} ($T_{AC}$) pseudo-phase transition
temperatures, and the smectic-\textrm{A} temperature range
($\Delta T_{A} = T_{NA} - T_{AC}$) all in kelvins and averaged
between heating and cooling scans. These are followed by similarly
averaged enthalpy ($\delta H_{NA}$) in J~g$^{-1}$ and specific
heat maximum $h_m \equiv \Delta C_p^{max}$ in J~K$^{-1}$~g$^{-1}$
values for the \textit{N}-Sm\textit{A} pseudo-phase transition.
The final column tabulates the specific heat step in
J~K$^{-1}$~g$^{-1}$ of the Sm\textit{A}-Sm\textit{C} phase
transition $\delta C_{AC}^{Step}$ averaged between heating and
cooling scans and taken as the value of the excess specific heat
$-6$~K below $T_{AC}$. }
\begin{ruledtabular}
\begin{tabular}{@{\extracolsep{22pt}}llllllll}
  $\rho_S$ & $l_o$ & $T_{NA}$ & $T_{AC}$ & $\Delta T_{A}$ & $\delta H_{NA}$ & $h_m$ & $\delta C_{AC}^{Step}$ \\
 \hline
  0      &$\infty$& 336.626  & 329.009  & 7.617  & 2.136  &  0.726  & 0.044 \\
  0.025  & 2636   & 336.073  & 328.836  & 7.237  & 1.985  &  0.541  & 0.043 \\
  0.041  & 1636   & 336.524  & 329.136  & 7.388  & 1.897  &  0.530  & 0.034 \\
  0.051  & 1303   & 336.315  & 329.124  & 7.191  & 1.780  &  0.377  & 0.025 \\
  0.078  &  859   & 336.519  & 329.527  & 6.992  & 1.834  &  0.487  & 0.032 \\
  0.105  &  636   & 336.676  & 329.210  & 7.466  & 1.952  &  0.444  & 0.027 \\
  0.161  &  414   & 336.573  & 329.456  & 7.117  & 1.573  &  0.302  & 0.014 \\
  0.220  &  303   & 336.865  & 329.915  & 6.950  & 1.561  &  0.284  & 0.013 \\
  0.347  &  192   & 336.505  & 329.621  & 6.884  & 1.404  &  0.231  & 0.003 \\
\end{tabular}
\end{ruledtabular}
\end{table*}

The nano-colloidal mixture of hydrophilic type-300 aerosil in
$\bar{8}$S5 was prepared following the solvent dispersion
procedure outlined in Refs.~\cite{Clegg03a,Clegg03b}. The
hydrophilic nature of the aerosils allows the silica particles to
weakly hydrogen bond to each other and form a gel in an organic
solvent. The specific surface area of type-300 aerosil is
$300$~m$^2$~g$^{-1}$~\cite{Degussa} and each aerosil sphere is
roughly $7$~nm in diameter. However, the basic free-floating
aerosil unit consists of a few of these spheres fused together
during the manufacturing process \cite{Germano98}. Each
$\bar{8}$S5+sil sample was created by mixing appropriate
quantities of liquid-crystal and aerosil together, then dissolving
the resulting mixture in spectroscopic grade (low water content)
acetone. The resulting solution was then dispersed using an
ultrasonic bath for about an hour. As the acetone evaporates from
the mixture, a fractal-like gel forms through diffusion-limited
aggregation. Crystallization of the LC host can severely disrupt
the gel structure and so care was taken to prevent any formation
of the solid phase of the liquid crystal during the experiments.

High-resolution ac calorimetry was performed using two home-built
calorimeters at WPI. The sample cell consisted of a silver
crimped-sealed envelope $\sim 10$~mm long, $\sim 5$~mm wide, and
$\sim 0.5$~mm thick (closely matching the dimensions of the
heater). After the sample was introduced into a cell having an
attached $120$-$\Omega$ strain-gauge heater and $1$-M$\Omega$
carbon-flake thermistor, a constant current was placed across the
heater to maintain the cell temperature well above $T_{IN}$. The
filled cell was then placed in an ultrasonic bath to remix the
sample. After remixing, the cell was mounted in the calorimeter,
the details of which have been described elsewhere \cite{Yao98}.
In the ac-mode, power is input to the cell as $P_{ac} e^{i\omega
t}$ resulting in temperature oscillations with amplitude $T_{ac}$
and a relative phase shift of $\varphi \equiv \Phi + \pi/2$, where
$\Phi$ is the absolute phase shift between $T_{ac}(\omega)$ and
the input power. The specific heat at a heating frequency $\omega$
using $C^\ast \equiv P_{ac}/\omega \left|T_{ac}\right|$ is given
by
\begin{equation}
  \label{eq:ReCp}
  C_p = \frac{[C_{filled}' - C_{empty}]}{m_{sample}} = \frac{C^\ast cos\varphi - C_{empty}}{m_{sample}} \;,
\end{equation}
\begin{equation}
  \label{eq:ImC}
  C_{filled}'' = C^\ast sin\varphi - \frac{1}{\omega R_e} \;,
\end{equation}
where $C_{filled}'$ and $C_{filled}''$ are the real and imaginary
components of the heat capacity, $C_{empty}$ is the heat capacity
of the cell and silica, $m_{sample}$ is the mass in grams of the
liquid crystal (the total mass of the $\bar{8}$S5+sil sample was
$\sim 20$~mg, which yielded $m_{sample}$ values in the range of
$13-20$~mg), and $R_e$ is the external thermal resistance linking
the cell and the bath (here, $\sim 200$~K~W$^{-1}$).
Equations~\eqref{eq:ReCp} and \eqref{eq:ImC} require a small
correction to account for the finite internal thermal resistance
compared to $R$, and this was applied to all samples studied here
\cite{Germano97}. Measurements were conducted at various
frequencies from $1$ to $500$~mHz in order to ensure the
applicability of Eqs.~\eqref{eq:ReCp} and \eqref{eq:ImC}. For all
results presented here, $C_{filled}^{''} \simeq 0$ through the
\textit{N}-Sm\textit{A} and Sm\textit{A}-Sm\textit{C} transitions,
which is expected for continuous phase transitions, and that $C_p$
was independent of $\omega$. All data presented here were taken at
a heating frequency of $\omega = 0.1473$~s$^{-1}$ (or $23.4$~mHz)
and a scanning rate of less than $\pm 100$~mK~h$^{-1}$, which
yielded static $C_p$ results. This equilibrium behavior relates
only to the pseudo-critical smectic fluctuations that are the
focus of this work. The glass-like behavior, that has been
observed in rheological studies in this frequency
range~\cite{Ranjini05}, is a characteristic of these systems away
from the pseudo-transition."

The bulk $\bar{8}$S5 and all $\bar{8}$S5+sil samples experienced
the same thermal history after mounting; six hours in the
isotropic phase to ensure homogeneous gelation, then a slow cool
into the smectic phase at $\sim 320$~K before beginning the first
detailed heating scan to $\sim 345$~K followed immediately by a
detailed cooling scan over the same range. Tests on bulk
$\bar{8}$S5 with various thermal histories in the isotropic phase
reveal an aging of the material. There are progressive shifts of
the transition temperatures downward with increased time at high
temperature. However, other than an increase in the rounding of
the $C_p$ peaks and the downward shift of the transition with
time, the critical character remains essentially unchanged.

\section{RESULTS}
\label{sec:results}

\begin{figure}
\includegraphics[scale=0.5]{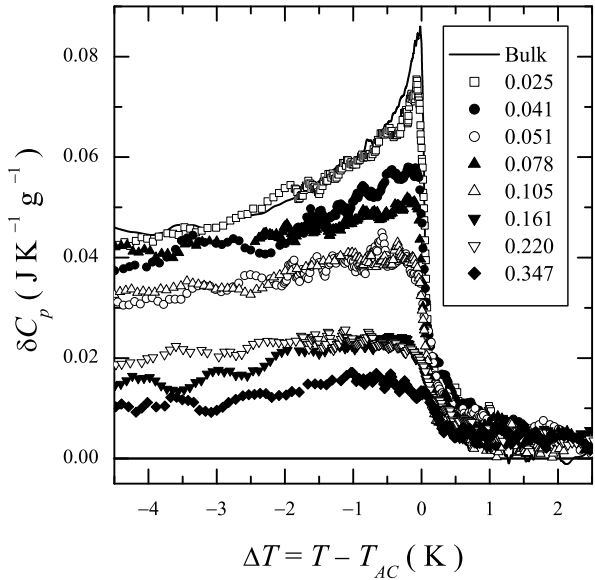}
\caption{ \label{dCPvsDTAC} Excess specific heat $\delta C_p$ of
the Sm\textit{A}-Sm\textit{C} phase transition from $+2.5$~K above
to $-4.5$~K below $T_{AC}$ for bulk $\bar{8}$S5 and
$\bar{8}$S5+sil samples. Inset lists the $\rho_S$ for each data
set shown. }
\end{figure}

The specific heat on cooling for a bulk sample of $\bar{8}$S5 is
shown in Fig.~\ref{CPvsT}. Clearly visible are the XY-like
\textit{N}-Sm\textit{A} phase transition at $336.71$~K and a
Landau mean-field Sm\textit{A}-Sm\textit{C} at $328.82$~K. These
are in good agreement with Schantz and Johnson~\cite{Schantz78}.
The excess specific heat due to the \textit{N}-Sm\textit{A}
transition $\Delta C_p(NA)$ for the bulk and LC+sil samples is
obtained by subtracting from the specific heat $C_p$ a linear
background
\begin{equation}
  \label{eq:DCna}
  \Delta C_p = C_p - C_p(\rm{background}) \;,
\end{equation}
where $C_p$(background) is shown as the dashed line in
Fig.~\ref{CPvsT} and represents the $C_p$ variation of the
low-temperature \textit{wing} of the \textit{I}-\textit{N}
transition. This expression is a valid representation of $\Delta
C_p(NA)$ for all $T > T_{AC}$ and the resulting $\Delta C_p(NA)$
for all samples studied are shown in Fig.~\ref{DCPvsDTNA} over a
$\pm 3.5$~K temperature range about $T_{NA}$. The fluctuation
dominated \textit{N}-Sm\textit{A} transition enthalpy is defined
as
\begin{equation}
  \label{eq:dHna}
  \delta H_{NA} = \int \Delta C_p(NA) dT \;,
\end{equation}
where consistent limits of the integration of $\pm 5$~K about
$T_{NA}$ were used for all samples. The specific heat contribution
of the Sm\textit{A}-Sm\textit{C} transition is obtained from
$\Delta C_p$ by subtracting the low-temperature heat capacity
\textit{wing} of the \textit{N}-Sm\textit{A} transition (see
Fig.~\ref{CPvsT}):
\begin{equation}
  \label{eq:DCac}
  \delta C_p(AC) = \Delta C_p - \Delta C^{wing}_p(NA) \;.
\end{equation}
The result for all samples studied are shown in
Fig.~\ref{dCPvsDTAC} over a range $+2$~K above to $-4.5$~K below
$T_{AC}$. The relevant thermal characteristics for both phase
transitions of $\bar{8}$S5+sils as a function of $\rho_S$ are
given in Table~\ref{tab:Cp-sum}.

As seen in Fig.~\ref{DCPvsDTNA}, the \textit{N}-Sm\textit{A} C$_p$
peak remains sharp for all aerosil densities up to the maximum
studied of $\rho_S = 0.347$ with no abrupt truncation marking a
transition from soft to stiff gel behavior seen in
8CB+sil~\cite{Germano98} and 8OCB+sil~\cite{Roshi04} systems. The
$\Delta C_p(NA)$ wings on both sides of the transition decrease
with increasing $\rho_S$ in contrast with that found for the 8CB
and 8OCB in aerosils, which found significant reduction of the
high-temperature $\Delta C_p(NA)$ wing \cite{Germano98}.
Qualitatively, the $\Delta C_p(NA)$ peak shape remains remarkably
stable with increasing $\rho_S$. For the $\bar{8}$S5+sil system,
the most striking effect of the aerosils is the strong decrease in
the heat capacity maximum at the transition, $h_m = \Delta
C_p^{max}(NA)$, and in the transition enthalpy $\delta H_{NA}$
with increasing $\rho_S$. In addition, the value of
$T_{NA}(\rho_S)$ also displays very little sensitivity to the
aerosils even when the aging of the $\bar{8}$S5 is taken into
account, in stark contrast to the non-monotonic downward shifts
found for nearly all other LC+sil systems \cite{Germano98}. See
Table~\ref{tab:Cp-sum}.

The effect of increasing aerosil concentration on the
Sm\textit{A}-Sm\textit{C} transition heat capacity appears
entirely at and below the transition $T_{AC}$. The $\delta
C_p(AC)$ peak smoothly evolves with increasing $\rho_S$ from the
bulk tricritical behavior to a simple heat capacity step as shown
in Fig.~\ref{dCPvsDTAC}. Also apparent is the systematic reduction
of the step magnitude with increasing $\rho_S$ as listed in
Table~\ref{tab:Cp-sum}. While there is some rounding of $\delta
C_p(AC)$ in the immediate vicinity of $T_{AC}$, there does not
appear any strong effect of the aerosil on the high-temperature
$\delta C_p$ tail. As with the \textit{N}-Sm\textit{A} transition,
even when aging of the bulk material is taken into account,
$T_{AC}(\rho_S)$ is not shifted downward with increasing aerosil
content but appears to shift slightly upward. The stability of
both $T_{NA}(\rho_S)$ and $T_{AC}(\rho_S)$ reported here is in
contrast to that found in the recent x-ray study that found large
non-monotonic transition temperature shifts with
$\rho_S$~\cite{Clegg03b}. However, the thermal history experienced
by these calorimetry samples was far less severe than that
experienced by the x-ray samples which were held at high
temperatures while the solvent evaporated prior to the
measurements. Thus, the results presented here should reflect a
consistent measure of the transition temperature shift with
$\rho_S$ for this system.

\subsection{The XY-like \textit{N}-Sm\textit{A} scaling analysis}

The shape of these experimental $\Delta C_p(NA)$ data as a
function of aerosil content is characterized by a power-law
form~\cite{Garland94} in terms of the reduced temperature $t =
\mid~\rm{T} - \rm{T}^\ast\mid~/~\rm{T}^\ast$ given by
\begin{equation}
  \label{eq:fitDCna}
  \Delta C_p(NA) = A^\pm t^{-\alpha} ( 1 + D^\pm t^{\Delta_1}) + B_c \;,
\end{equation}
where the critical behavior as a function of reduced temperature
$t$ is characterized by an exponent $\alpha$, amplitudes $A^\pm$
above and below the transition, a critical background term $B_c$,
and corrections-to-scaling terms having an amplitude $D^\pm$ and
exponent $\Delta_1 \simeq 0.5$. An increasing temperature gap of
excluded data about the $\Delta C_p(NA)$ peak with increasing
$\rho_S$ was required to perform the nonlinear least-squares
fitting. These excluded data were identified by strong deviations
in the residuals near $T^\ast$ and represent rounded $\Delta
C_P(NA)$ values not describable by the power-law given in
Eq.~\eqref{eq:fitDCna}. These fit results for the sets of heating
and cooling scans of bulk $\bar{8}$S5 and $\bar{8}$S5+sil samples
are presented in Tables~\ref{tab:NA-fitH} and~\ref{tab:NA-fitC},
respectively. In addition to fits with the effective critical
exponent $\alpha_{eff}$ as a free parameter, fits were also
performed with $\alpha_{eff}$ fixed to the 3D-XY value of
$\alpha_{\rm{XY}} = -0.013$. The bulk $\bar{8}$S5 fit results are
of good quality for both heating and cooling and yield a very
small negative $\alpha_{eff}$ that is not significantly different
from fixing $\alpha_{eff}$ to $\alpha_{\rm{XY}} = -0.013$ as
indicated by $\chi_\nu^2$~\cite{noteCHI2}. This indicates that the
specific heat of the \textit{N}-Sm\textit{A} transition for the
bulk LC behaves essentially as a clean XY transition consistent
with the literature~\cite{Schantz78,Marinelli96}. However, it
should be noted that the critical exponents for the parallel
correlation length ($\nu_\parallel$) and for the smectic
susceptibility ($\gamma$) do not have XY values~\cite{Garland94}
and that the \textit{N}-Sm\textit{A} transition for bulk
$\bar{8}$S5 exists in a complex crossover regime.

\begin{table*}
\caption{ \label{tab:NA-fitH} Heating scan summary of the results
of fitting Eq.~\eqref{eq:fitDCna} to the excess specific heat peak
$\Delta C_p$ of the \textit{N}-Sm\textit{A} phase transition on
$\bar{8}$S5+sil samples. The transition temperature $T_C$ is given
in kelvins, while the parameters $B_C$ and $A^\pm$ are given in
J~K$^{-1}$~g$^{-1}$. The parameters $D^\pm$ are dimensionless. All
scans were fit from $t_{max} = 10^{-2}$ to $\pm t_{min}$. All
parameters were free to vary in the fit except when the exponent
was fixed to $\alpha_{\textrm{XY}} = -0.013$ (denoted by the
square brackets).}
\begin{ruledtabular}
\begin{tabular}{@{\extracolsep{11pt}}lllrrrrrll}
 $\rho_S$ & $T_C$ & $\alpha_{eff}$ & $B_C$ & $A^+$ & $A^-$ & $D^+$ & $D^-$ & $t_{min}\times10^{-5}$ & $\chi_\nu^2$ \\
 \hline \\
  0     & 335.220 &-7$\times10^{-5}$& 1023.600 &-1023.760 &-1023.630 & 0.0002 & 0.0022 & +1.80/-12.3 & 1.266 \\
        & 335.222 & [-0.013]        &    6.053 &   -6.260 &   -6.122 & 0.0026 & 0.3712 &             & 1.319 \\
  0.025 & 334.712 &-1$\times10^{-4}$&  536.893 & -536.971 & -536.845 & 0.0007 & 0.0045 & +3.50/-29.9 & 1.073 \\
        & 334.711 & [-0.013]        &    4.757 &   -4.880 &   -4.741 & 0.0450 & 0.5234 &             & 1.027 \\
  0.041 & 335.111 &-1$\times10^{-4}$&  383.615 & -383.643 & -383.536 & 0.0016 & 0.0060 & +2.56/-14.2 & 1.287 \\
        & 335.110 & [-0.013]        &    4.124 &   -4.194 &   -4.076 & 0.1117 & 0.5715 &             & 1.307 \\
  0.051 & 334.938 &-2$\times10^{-4}$&  154.089 & -154.083 & -154.025 & 0.0045 & 0.0086 & +11.4/-49.2 & 1.134 \\
        & 334.937 & [-0.013]        &    3.228 &   -3.249 &   -3.186 & 0.1954 & 0.4153 &             & 1.137 \\
  0.078 & 335.123 &-1$\times10^{-4}$&  376.628 & -376.688 & -376.614 & 0.0010 & 0.0036 & +3.08/-14.9 & 1.163 \\
        & 335.123 & [-0.013]        &    4.049 &   -4.146 &   -4.064 & 0.0628 & 0.3348 &             & 1.183 \\
  0.105 & 335.297 &-1$\times10^{-4}$&  259.368 & -259.267 & -259.171 & 0.0049 & 0.0095 & +3.63/-19.3 & 1.076 \\
        & 335.297 & [-0.013]        &    2.767 &   -2.698 &   -2.593 & 0.4513 & 0.9716 &             & 1.084 \\
  0.161 & 335.162 &-7$\times10^{-4}$&   18.744 &  -18.608 &  -18.580 & 0.0655 & 0.0615 & +5.67/-42.3 & 1.036 \\
        & 335.161 & [-0.013]        &    1.335 &   -1.212 &   -1.181 & 1.0162 & 0.9862 &             & 1.037 \\
  0.220 & 335.498 &-4$\times10^{-4}$&   10.314 &  -10.099 &  -10.048 & 0.1651 & 0.1875 & +3.59/-57.5 & 1.004 \\
        & 335.496 & [-0.013]        &    0.630 &   -0.428 &   -0.373 & 3.9745 & 5.2388 &             & 1.004 \\
  0.347 & 335.145 &-6$\times10^{-4}$&    6.892 &   -6.724 &   -6.703 & 0.1839 & 0.1400 & +0.27/-85.1 & 1.056 \\
        & 335.144 & [-0.013]        &    0.573 &   -0.414 &   -0.391 & 3.0646 & 2.4651 &             & 1.056 \\
  \\
\end{tabular}
\end{ruledtabular}
\end{table*}

\begin{table*}
\caption{ \label{tab:NA-fitC} Cooling scan summary of the results
of fitting Eq.~\eqref{eq:fitDCna} to the excess specific heat peak
$\Delta C_p$ of the \textit{N}-Sm\textit{A} phase transition on
$\bar{8}$S5+sil samples. The labels are the same as those used in
Table~\ref{tab:NA-fitH} as well as the uncertainties of the
adjustable fit parameters. }
\begin{ruledtabular}
\begin{tabular}{@{\extracolsep{11pt}}lllrrrrrll}
 $\rho_S$ & $T_C$ & $\alpha_{eff}$ & $B_C$ & $A^+$ & $A^-$ & $D^+$ & $D^-$ & $t_{min} \times 10^{-5}$ & $\chi_\nu^2$ \\
 \hline \\
  0     & 335.198 &-7$\times10^{-5}$& 1047.600 &-1047.800 &-1047.670 &-0.0001 & 0.0019 & +3.42/-9.91 & 1.463 \\
        & 335.197 & [-0.013]        &    6.565 &   -6.826 &   -6.688 &-0.0499 & 0.2870 &             & 1.532 \\
  0.025 & 334.632 &-1$\times10^{-4}$&  474.449 & -474.466 & -474.327 & 0.0012 & 0.0060 & +7.86/-32.8 & 1.159 \\
        & 334.630 & [-0.013]        &    4.422 &   -4.484 &   -4.334 & 0.0956 & 0.6721 &             & 1.176 \\
  0.041 & 335.108 &-3$\times10^{-4}$&  141.506 & -141.494 & -141.395 & 0.0056 & 0.0167 & +4.37/-23.5 & 1.203 \\
        & 335.108 & [-0.013]        &    3.844 &   -3.868 &   -3.760 & 0.1782 & 0.6412 &             & 1.207 \\
  0.051 & 334.845 &-1$\times10^{-4}$&  216.789 & -216.692 & -216.619 & 0.0047 & 0.0091 & +11.0/-36.4 & 1.134 \\
        & 334.843 & [-0.013]        &    2.611 &   -2.541 &   -2.462 & 0.3824 & 0.8130 &             & 1.138 \\
  0.078 & 335.057 &-2$\times10^{-4}$&  195.069 & -195.050 & -195.018 & 0.0050 & 0.0069 & +32.6/-32.7 & 1.188 \\
        & 335.055 & [-0.013]        &    3.800 &   -3.809 &   -3.775 & 0.2461 & 0.3545 &             & 1.192 \\
  0.105 & 335.179 &  -0.0043        &   11.848 &  -11.888 &  -11.823 & 0.0425 & 0.1152 & +37.0/-68.2 & 1.145 \\
        & 335.178 & [-0.013]        &    4.239 &   -4.301 &   -4.232 & 0.1061 & 0.3208 &             & 1.145 \\
  0.161 & 335.123 &-4$\times10^{-4}$&   48.578 &  -48.455 &  -48.427 & 0.0192 & 0.0211 & +16.4/-39.1 & 1.151 \\
        & 335.121 & [-0.013]        &    1.647 &   -1.538 &   -1.508 & 0.6025 & 0.6853 &             & 1.152 \\
  0.220 & 335.458 &-3$\times10^{-4}$&   33.646 &  -33.468 &  -33.415 & 0.0363 & 0.0481 & +19.7/-99.6 & 1.032 \\
        & 335.455 & [-0.013]        &    1.205 &   -1.043 &   -0.985 & 1.1714 & 1.6754 &             & 1.032 \\
  0.347 & 335.14  &-3$\times10^{-4}$&    9.953 &   -9.734 &   -9.706 & 0.1367 & 0.1229 & +5.65/-74.2 & 1.129 \\
        & 335.14  & [-0.013]        &    0.332 &   -0.119 &   -0.088 &11.6406 &14.1668 &             & 1.129 \\
  \\
\end{tabular}
\end{ruledtabular}
\end{table*}

As $\rho_S$ increases for the $\bar{8}$S5+sil samples, the fits
for both heating and cooling are all of good quality even up to
the highest density studied. In general, the adjustable parameters
have approximately a $10\%$ uncertainty in magnitude for the bulk
and the two lowest density aerosil samples, except for $T_c$ which
has an uncertainty of approximately $\pm 5$~mK for all fits. This
uncertainty grows for the higher density samples because an
increasing temperature gap about the $\Delta C_p$ peak of data is
excluded from the fits due to rounding. These fits show that the
critical character described by the exponent $\alpha_{eff}$ of the
\textit{N}-Sm\textit{A} transition remains unchanged in the
$\bar{8}$S5+sil samples for all $\rho_S$, even above the $\rho_S =
0.1$ where previous studies of other LC+sil samples found no
critical behavior~\cite{Germano98,Germano04}. It is apparent from
the fits that the increasing temperature gap ($\pm t_{min}$) is
dominated by rounding on the low-temperature side of the $\delta
C_p(NA)$ peak. This increasing temperature gap for the fits has
the effect of causing the increase in the magnitudes of the
correction-to-scaling coefficients $D^\pm$ although they remain
well behaved. Fixing the ratio $D^{+}/D^{-} = 1$ or setting $D^{+}
= D^{-} = 0$ resulted in significantly poorer fits for all
samples. It is also evident is that the critical amplitudes above
and below the transition, $A^\pm$ systematically decrease with
increasing $\rho_S$ although the ratio $A^+/A^-$ remains
unchanged.

The scaling analysis for the \textit{N}-Sm\textit{A} transition in
LC+aerosil systems follows that of reference~\cite{Germano03} by
using the correlation length power-law to equate the cut-off
length scale (maximum correlation length) $\xi_M$ to a minimum
reduced temperature. This is closely related to a finite-size
effect, however, there are no true pores and the LC filled voids
are highly interconnected. The correlation length is given by
\begin{equation}
  \label{eq:appXIPARA}
  \xi_\parallel = \xi_{\parallel o} t^{-\nu_\parallel} \;.
\end{equation}
Since Eq.~\eqref{eq:appXIPARA} is defined only for $T > T^\ast$,
the finite-scale induced rounding of the transition is estimated
in terms of the minimum reduced temperature on the
high-temperature side of the transition $t_m^+$, by the form
\begin{equation}
  \label{eq:dTnaFSS}
  \delta T^\ast / T^\ast \approx 2t_m^+ =
  2\left(\frac{\xi_M}{\xi_{\parallel o}}\right)^{-1/\nu_\parallel}
  = 2\left(\frac{\xi_{\parallel o}}{2n}
  a\rho_S\right)^{1/\nu_\parallel} \;,
\end{equation}
where the cut-off length scale is written in terms of $n$, the
number of mean-void lengths $l_o$. The heat capacity maximum at
the transition $h_m$ is given by substituting $t_m^+$ into
Eq.~\eqref{eq:DCna} and is also explicitly defined for the $T >
T^\ast$. The explicit form for $h_m$ is
\begin{equation}
  \label{eq:DCnaFSS}
  h_m = A^+ \left(\frac{\xi_M}{\xi_{\parallel o}}\right)^{\alpha / \nu_\parallel}
  \left( 1 + D^+ \left(\frac{\xi_M}{\xi_{\parallel o}}\right)^{-\Delta_1 / \nu_\parallel} \right)
  + B_c \;.
\end{equation}
Finally, the transition enthalpy $\delta H_{NA}$ resulting from
this scaling analysis is determined by using $t_m^+$ to truncate
the integration of Eq.~\eqref{eq:DCna} both above and below the
transition $T^\ast$.

\begin{figure}
\includegraphics[scale=0.5]{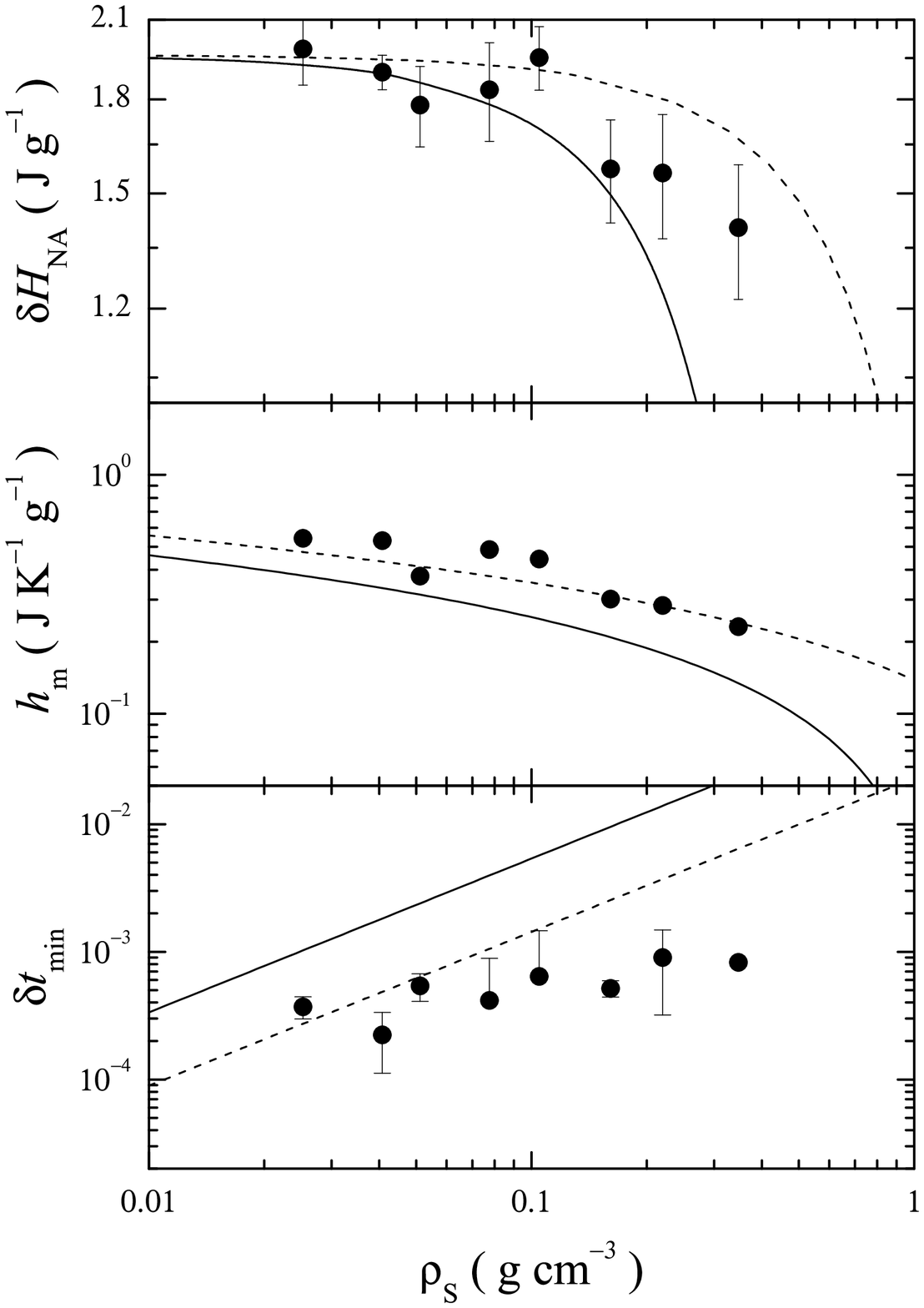}
\caption{ \label{NAfss} Scaling analysis of the
\textit{N}-Sm\textit{A} phase transition in $\bar{8}$S5+sil (solid
circles) on log-log scales. Top Panel: Transition enthalpy $\delta
H_{NA}$. Middle Panel: Excess specific heat maximum $h_m = \Delta
C_{NA}^{max}$. Bottom Panel: Transition rounding in reduced
temperature $\delta t_{min} = |t_{min}^-| + |t_{min}^+|$. The bulk
scaling predictions are given by Eqs.~\eqref{eq:DCnaFSS} and
\eqref{eq:dTnaFSS} where the mean-void length $l_o$ is used as the
cut-off length scale (solid lines) or $3l_o$ (dashed lines). }
\end{figure}

Plotted on Fig.~\ref{NAfss} are the scaling trends for the
\textit{N}-Sm\textit{A} transition of $\bar{8}$S5+sil samples
using the \textit{bulk} \textit{N}-Sm\textit{A} $\bar{8}$S5
critical parameters. Two choices for the cut-off length scale are
shown. The first choice for the cutoff correlation length $\xi_M$
uses the mean distance between silica surfaces (mean void size)
$l_o = 2 / a \rho_S$, where $a$ is the specific surface
area~\cite{Germano04}. The second choice allows $\xi_M$ to vary as
some multiple of $l_o$, i.e. $\xi_M = nl_o$. The results for $h_m$
of the $\bar{8}$S5+sil samples are in very good agreement using
$\xi_M \equiv 3l_o$, which more closely matches the measured
saturated smectic correlation length~\cite{Clegg03b}. However,
$\delta T^\ast / T^\ast$ appears \textit{sharper} and $\delta
H_{NA}$ \textit{smaller} for $\rho_S > 0.1$ than predicted by this
analysis.

\subsection{The Landau mean-field Sm\textit{A}-Sm\textit{C} scaling analysis}

\begin{figure}
\includegraphics[scale=0.5]{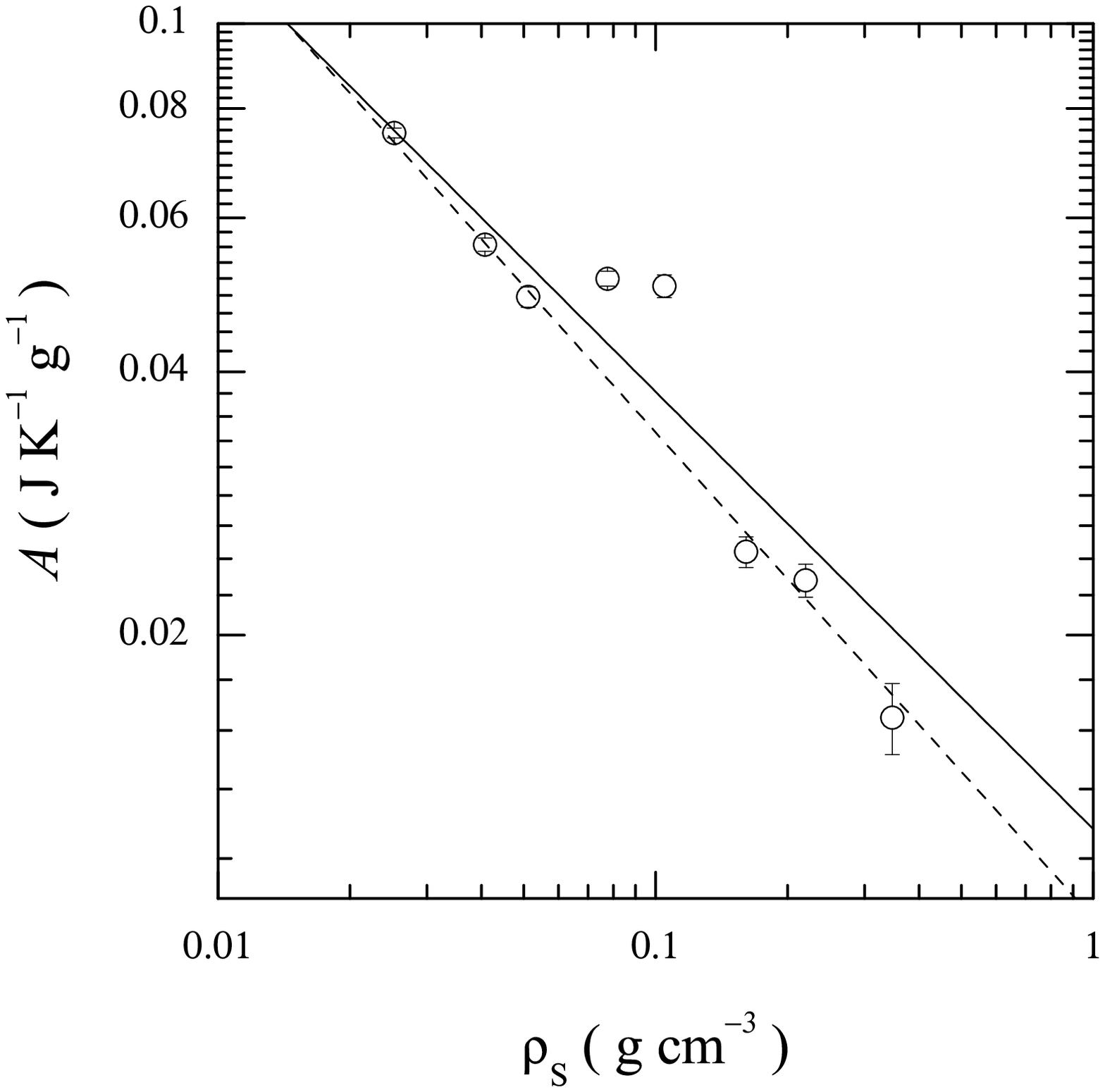}
\caption{ \label{AvsRHOS} Scaling plot of $A$ ($\delta C_p$ at $T
= T_{AC}$), obtain from fitting the excess specific heat of the
Sm\textit{A}-Sm\textit{C} phase transition to
Eq.~\eqref{eq:fitDCac}, against $\rho_S$. The solid line has a
slope of $-0.5$ while the dashed line a slope of $-0.56$. }
\end{figure}

The excess specific heat due to the Sm\textit{A}-Sm\textit{C}
phase transition is shown in Fig.~\ref{dCPvsDTAC} for bulk
$\bar{8}$S5 and $\bar{8}$S5+sil samples over a temperature range
down to $T_{AC} - 4.5$~K. The bulk Sm\textit{A}-Sm\textit{C}
transition is well described by an extended Landau
theory~\cite{Schantz78,Meichle83} given by
\begin{equation}
  \label{eq:fitDCac}
  \delta C_p(AC) =
  \left\{
  \begin{array}{l}
    0 \\\\
    A \frac{T}{T_c}\left(\frac{T_m - T_c}{T_m - T}\right)^{0.5}
  \end{array}
  \begin{array}{r}
    \textrm{for}~T > T_c \\\\
    \textrm{for}~T < T_c
  \end{array}
  \right.\;
\end{equation}
where $A$ is the $\delta C_p(AC)$ maximum at the transition $T_c$
and $T_m$ is the upper stability limit of the Sm\textit{C} phase.
Results from fitting Eq.~\eqref{eq:fitDCac} to the excess specific
heat of the Sm\textit{A}-Sm\textit{C} phase transition $\delta
C_p$ for bulk $\bar{8}$S5 and $\bar{8}$S5+sil samples over the
temperatures $T < T_{AC}$ are tabulated in Table~\ref{tab:AC-fit}.
These results are the average of heating and cooling scan fits.
The stability limit of the Sm\textit{C} phase appears to shift
upwards from the bulk value by approximately $1.6$~K with
increasing $\rho_S$. Although intriguing and indicating a
stabilization of the Sm\textit{C} phase with increased $\rho_S$,
given the aging of this particular LC, it is not clear how robust
is this result. The fit coefficient $A$ of Eq.~\eqref{eq:fitDCac}
appears to smoothly decrease with increasing $\rho_S$. This is
consistent with the behavior of the Sm\textit{A}-Sm\textit{C}
transition of 7O.4 in hydrophobic aerosil~\cite{Haga97b}. The
parameter $A$ is also a measure of the tricritical nature of the
Sm\textit{A}-Sm\textit{C} transition. A log-log plot shown in
Fig.~\ref{AvsRHOS} of $A$ versus $\rho_S$ reveals a power-law
scaling given by $A \propto \rho_S^{-0.5}$.

\begin{table*}
\caption{ \label{tab:AC-fit} Summary of the results of fitting
Eq.~\eqref{eq:fitDCac} to the excess specific heat peak $\delta
C_p$ of the Sm\textit{A}-Sm\textit{C} phase transition for
$\bar{8}$S5+sil samples. The fit results have been averaged
between heating and cooling scans and were done from $\sim -7$~K
below $T_{AC}$ to a point slightly below the rounded $\delta C_p$
peak. The transition temperature $T_C$ and the Sm\textit{C}
stability limit $T_m$ are given in kelvins while the coefficient
$A$ is given in J~K$^{-1}$~g$^{-1}$. No uncertainty is quoted for
$T_C$ as it was fixed to this final value for the last fit
iteration. }
\begin{ruledtabular}
\begin{tabular}{@{\extracolsep{22pt}}ccccc}
  $\rho_S$ & $T_C$ & $T_m$ & $A$ & $\chi_\nu^2$ \\
 \hline
  0      & 328.957  & 329.26 $\pm$ 0.04  &  0.108 $\pm$ 0.001  & 1.187 \\
  0.025  & 328.706  & 330.32 $\pm$ 0.38  &  0.075 $\pm$ 0.002  & 1.208 \\
  0.041  & 329.090  & 330.61 $\pm$ 0.36  &  0.056 $\pm$ 0.002  & 1.328 \\
  0.051  & 329.010  & 332.23 $\pm$ 0.31  &  0.049 $\pm$ 0.003  & 1.323 \\
  0.078  & 329.495  & 330.49 $\pm$ 0.25  &  0.051 $\pm$ 0.002  & 1.258 \\
  0.105  & 329.250  & 331.08 $\pm$ 0.40  &  0.050 $\pm$ 0.010  & 2.317 \\
  0.161  & 329.390  & 329.97 $\pm$ 0.44  &  0.025 $\pm$ 0.002  & 1.160 \\
  0.220  & 329.860  & 331.41 $\pm$ 0.19  &  0.023 $\pm$ 0.003  & 1.206 \\
  0.347  & 329.414  & 330.37 $\pm$ 0.47  &  0.016 $\pm$ 0.002  & 1.211 \\
\end{tabular}
\end{ruledtabular}
\end{table*}

\section{DISCUSSION}
\label{sec:disc}

Results have been presented from a series of high-resolution
ac-calorimetric experiments on $\bar{8}$S5+sil dispersions through
the \textit{N}-Sm\textit{A} and Sm\textit{A}-Sm\textit{C} phase
transitions as a function of aerosil density. In bulk $\bar{8}$S5,
the \textit{N}-Sm\textit{A} is a nearly clean XY-like transition
with an $\alpha \lesssim 0$ while the Sm\textit{A}-Sm\textit{C}
transition is mean-field with a Landau tricritical character. Our
bulk measurements are fully consistent with these behaviors. The
introduction of QRD in the $\bar{8}$S5+sil system allows for the
isolation of random-field, finite-size-like, and elastic strain
effects at a pure XY and mean-field transition.

\begin{figure}
\includegraphics[scale=0.5]{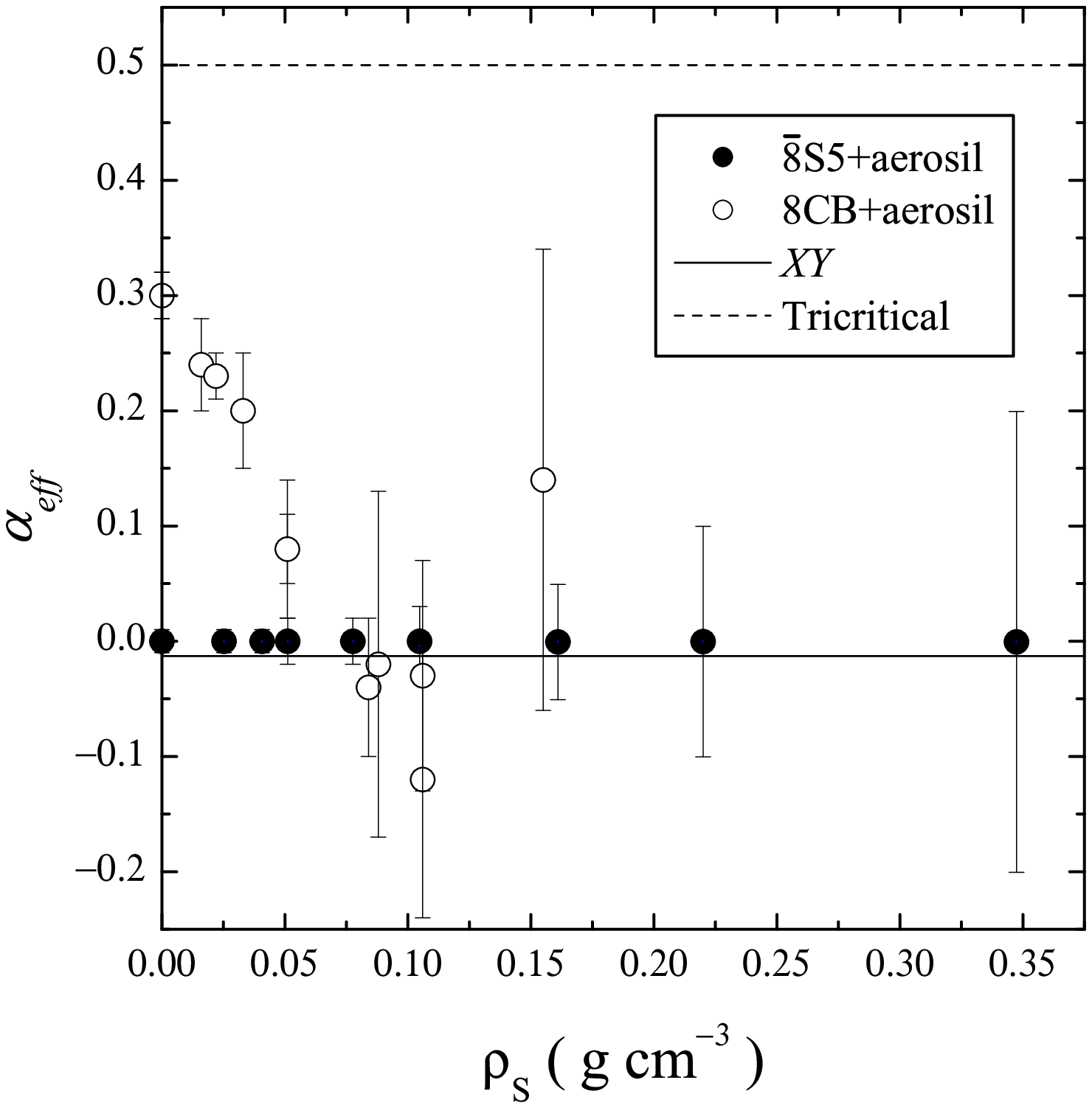}
\caption{ \label{ALPHAcompare} Comparison of effective specific
heat critical exponent $\alpha_{eff}$ obtained from fitting
Eq.~\eqref{eq:fitDCna} to the \textit{N}-Sm\textit{A} phase
transition data for $\bar{8}$S5+sil (solid circles) and
8CB+aerosil (open circles) samples. }
\end{figure}

The \textit{N}-Sm\textit{A} transition transition temperature does
not exhibit a sensitivity to the aerosil (and in fact increases
slightly) in stark contrast to that seen in nearly all other
LC+sil studies. Although the $\Delta C_p(NA)$ decreases uniformly
in size with increasing $\rho_S$, it remains sharp and is well
characterized over the whole range of aerosil concentration. The
power-law fits reveal that quasi-critical behavior is preserved
and that the exponent $\alpha_{eff}$ remains slightly negative and
constant for all $\rho_S$ studied. See Fig.~\ref{ALPHAcompare}.
The systematic decrease of the coefficients $A^\pm$ but nearly
constant ratio $A^+/A^-$ is a reflection of the uniform decrease
in the size of $\Delta C_p(NA)$ (as well as $\delta H_{NA}$). This
analysis is consistent with the view that the aerosil gel
effectively increases the nematic range. That is the nematic
susceptibility decreases with increasing $\rho_S$ and this is
reflected in a substantial change in the pseudo-critical
properties for cyanobiphenyl-aerosil systems. For $\bar{8}$S5+sil
where the bulk material already has a long nematic range, little
change is observed.

Because of the good quality power-law fits available in this
study, a detailed scaling analysis was performed and compared to
the heat capacity maximum $h_m = \Delta C_p^{max}(NA)$, transition
enthalpy $\delta H_{NA}$, and transition temperature rounding
$\delta T^\ast/T^\ast$ of the \textit{N}-Sm\textit{A} phase
transition. This analysis uses the bulk parameters and an
adjustable cutoff length scale $\xi_M$ and found excellent
agreement for $h_m$ using $\xi_M(\bar{8}\textrm{S5+sil}) = 3l_o$
but this does not completely describe the behavior of $\delta
H_{NA}$ nor $\delta T^\ast/T^\ast$, especially for $\rho_S > 0.1$.
Interestingly, when this scaling analysis is applied to the
8CB+sil and 8OCB+sil systems, equally good modelling of $h_m$ is
obtained for $\xi_M(\textrm{8CB+sil}) = l_o$ and
$\xi_M(\textrm{8OCB+sil}) = 1.5l_o$. It is surprising that such
excellent agreement is obtained over the whole range of $\rho_S$
explored for these three LCs despite the apparent violation of the
classic expectation of finite-size scaling (that is, the
truncation of the bulk $\Delta C_p$ behavior at the cut-off
length-scale). Also, the smectic correlation lengths measured for
these systems are much larger than this cutoff length
scale~\cite{Clegg05}. This makes the geometric interpretation of
the cut-off length puzzling but ultimately connected to the
strength of the disorder.

For the mean-field Sm\textit{A}-Sm\textit{C} phase transition, the
transition temperature also exhibits very weak sensitivity to the
presence of aerosil (even increasing slightly rather than
decreasing). This is also in stark contrast to that seen for the
Sm\textit{A}-Sm\textit{C} transition in 7O.4+sil and $\bar{8}$S5
in aerogel~\cite{Germano97} as well as the
Sm\textit{A}-Sm\textit{C}$^\ast$ transition of CE8+sil. The heat
capacity associated with the Sm\textit{A}-Sm\textit{C} transition
in $\bar{8}$S5+sil presented in this work exhibit a systematic
evolution from a Landau tricritical peak to a simple mean-field
step with very little smearing observed for $T > T_{AC}$, and then
only at the highest $\rho_S$. Good quality fits were made using an
extended Landau form that found the upper stability temperature
increasing with $\rho_S$ consistent with the shift of the observed
transition. In addition, the coefficient $A$, representing the
heat capacity maximum at the Sm\textit{A}-Sm\textit{C} transition,
exhibits a scaling with the aerosil conjugate density as $A
\propto \rho_S^{-0.5}$, see Fig.~\ref{AvsRHOS}. It is clear that
this transition, while evolving in a systematic fashion, remains
mean-field over the entire range of $\rho_S$ studied. This may be
a reflection of the Sm\textit{A}-Sm\textit{C} transition behaving
as if it were at its upper critical dimension moderating the
effect of the QRD produced by the aerosil gel.

In this study on $\bar{8}$S5+sil, it appears that the random-field
QRD of the aerosil gel plays a relatively weak role at the
\textit{N}-Sm\textit{A} transition (due to the non-polar molecules
and the long nematic range) nor at the Sm\textit{A}-Sm\textit{C}
transition (due to either its proximity to its upper critical
dimension or that the QRD does not couple linearly to the
Sm\textit{C} order parameter and so, does not provide a
random-field interaction). It is clear that a special length scale
is present with $\xi_M > l_o$ but it does not appear to play a
leading role for the Sm\textit{A}-Sm\textit{C} transition. How,
then, does one understand the evolution of this transition with
increasing QRD of aerosil and in light of the other LC+sil and
LC+aerogel results? One possibility is that the aerosil gel, due
to its flexibility, is closer to thermodynamic equilibrium with
the host LC and so causes the LC+sil to behave as a stiffer LC as
well as provide QRD. The aerosil gel has been shown to exhibit
dynamics coupled to the host liquid crystal~\cite{Retsch02} and
recent work has followed its quenching as aerosil density
increases~\cite{Ranjini05,Roshi05}. This increase in the effective
microscopic elastic stiffness of the LC would be similar to the
engineered stiffening of polymer composite materials. A
consequence of the stiffening of the nematic phase in LC+sil
systems has already been discussed
previously~\cite{Germano98,Germano03,Germano04,Clegg05} in terms
of the decrease in the nematic susceptibility with increasing
$\rho_S$ to explain the crossover from tricritical to XY behavior
for the \textit{N}-Sm\textit{A} transition. The present view is to
extend this concept to the general stiffening of the LC within the
LC+sil.

In the case of the Sm\textit{A}-Sm\textit{C} transition the
quenched disorder is also changing the character of the
transition. With increasing aerosil density, the peak at the
transition is being suppressed in favor of a step-like behavior.
This indicates that the transition is being driven away from
tricriticality toward mean-field character alone. The proximity to
tricriticality is controlled by the layer compression elastic
constant, $B$, in the modified fourth-order free energy
coefficient $b = b_0 - 2 \lambda^2 / B$, where $b_0$ is the
original value of this coefficient and $\lambda$ is the strength
of the coupling between the tilt angle and the layer
compression~\cite{Gennes93}. The mean-field phase transition
behavior and the decreasing size of the step in the heat capacity
is indicative of a hardening of the layer compression elastic
constant $B$.

This effect would serve to explain the evolution of the
Sm\textit{A}-Sm\textit{C} transition from Landau tricritical to
step in terms of a stiffening of the smectic layer compression
modulus $B$ and account for the general stability of the
transition temperatures. Note that for $\bar{8}$S5, the nematic
and smectic phases are almost fully decoupled in the bulk while
there remains significant coupling in the 8CB and 8OCB LCs. It is
the latter two LCs that exhibit the strong non-monotonic
transition temperature shifts downward with $\rho_S$ and so may be
reflecting QRD effects at the \textit{I}-\textit{N}
transition~\cite{Caggioni04}. It would be expected that twisted
(or chiral) phases would be more disordered and smeared due to
this general LC stiffening as seen in a recent
Sm\textit{A}-Sm\textit{C}$^\ast$ in aerosil
study~\cite{Kutnjak05}. Also, if the silica gel is too rigid, then
this LC stiffening effect is supplanted by the strong quenched
disorder as seen in LC in aerogel and in high density aerosil gel.

Finally, there seems to be an apparent inconsistency between the
experimental results of the x-ray study~\cite{Clegg03b} and this
calorimetry study on $\bar{8}$S5+sil. In the x-ray work, it is
clear that the temperature dependence of the tilt angle of the
Sm\textit{C} phase is insensitive to $\rho_S$, strongly suggesting
that the order-parameter $\psi$ for the Sm\textit{C} phase remains
unchanged from bulk for all $\bar{8}$S5+sil samples. However, if
the system remains mean-field, then the relationship $C_p \propto
\partial \psi^2 / \partial T$ should hold. In this case, the dramatic variation of
$\delta C_p(AC)$ with $\rho_S$ suggests that the Sm\textit{C}
order-parameter should be strongly $\rho_S$ dependent. Comparison
of the $\delta C_p$ data in Fig.~\ref{dCPvsDTAC} (one data point
every $0.1$~K) and the tilt angle data in Fig.~6 in
ref.~\cite{Clegg03b} (one data point every $\approx 1$~K) suggest
that the situation is more ambiguous. A suppression of the
tricritical character may be consistent with the more sparse x-ray
data (see note [21] in ref.~\cite{Clegg03b}).

Clearly, there is an important need for theoretical efforts at
understanding quenched random disorder on the
Sm\textit{A}-Sm\textit{C} transition. The beginnings of such a
theory, although for the anisotropic aerogel disordered
Sm\textit{A}-Sm\textit{C} transition~\cite{Chen05}, are emerging
but much is left to be done. A coherent framework must account for
the apparent connection with elasticity of both the host material
and the disorder inducing gel. It must also provide for an
interpretation of the cut-off length scale that seems to play an
important role for the \textit{N}-Sm\textit{A} as well as the
observed scaling of the $\Delta C_p(AC)$ peak with $\rho_S$.
Additional x-ray and calorimetric experimental work with
anisotropically aligned Sm\textit{A}-Sm\textit{C} as well as
Sm\textit{A}-Sm\textit{C}* in aerosil gel systems are also needed
and would provide important data as to the role the polar nature
of the LC and the local aerosil-LC interaction.

\begin{acknowledgments}

The authors wish to thank N. A. Clark, C. W. Garland, R. Leheny,
and T. Bellini for many helpful and useful discussions. P.S.C. and
R.J.B. wish to thank the Natural Science and Engineering Research
Council of Canada while staying in Toronto. The funding in
Edinburgh was provided by the EPSRC (Grant GR/S10377/01). The work
at WPI was supported by the NSF under the NSF-CAREER award
DMR-0092786.

\end{acknowledgments}





\begin{thebibliography}{38}
\bibitem[{not()}]{noteCHI2}
\bibinfo{note}{The reduced chi-squared parameter $\chi_\nu^2 =
  \frac{1}{\nu}\chi^2 = \frac{1}{\nu}\sum\frac{(y_i - f(x_i))^2}{\sigma_i^2}$,
  where $\nu = N - n$ is the degrees of freedom fitting $N$ data with $n$
  parameters in a model $f(x_i)$, is the mean-sqaured deviations normalized to
  measurement errors $\sigma_i$ in $y_i$.}
\end{thebibliography}

\end{document}